
%
%
\magnification=1200
\baselineskip=14truept
\vsize=8.75truein
\hsize=6truein
\font\smrm =cmr10 at 10truept
\vbox{\vskip70truept}
\centerline{\bf DEVELOPMENTS IN THE THEORY OF THE }
\centerline{\bf QUANTUM HALL EFFECT}
\vskip56truept
\centerline{BERTRAND I. HALPERIN}
\centerline{Physics Department, Harvard University}
\centerline{Cambridge, MA 02139, U.S.A.}
\vskip28truept
\centerline{ABSTRACT}
\vskip28truept

\noindent
\hskip.5truein\vbox{\hsize5truein\baselineskip=12truept{\noindent\smrm
The past few years have
produced major advances in our understanding of the quantum Hall
effects---quantized
and unquantized. Theories based on a mathematical transformation, where the
electrons are replaced by a set of fermions interacting with a Chern-Simons
gauge
field, have been useful for explaining and predicting observations at
even-denominator filling fractions where quantized Hall plateaus are not
observed, as well as for giving new insight into the most prominent
fractional quantized Hall states at odd-denominator fractions. Other
theoretical approaches have led to important advances in our understanding
of edge-excitations for systems in a fractional quantized Hall state,
of phases and phase transitions in bilayer systems, of tunneling phenomena in
the quantum Hall regimes, and of disorder-induced transitions between
``neighboring'' quantum Hall plateaus. Some highlights of these developments
will
be reviewed.}
}
\vskip14truept

\noindent
{\bf 1. Introduction}
\vskip14truept
During the past few years, there has been very substantial progress, both
theoretical and experimental, in our understanding of the behavior of electrons
in a
partially-filled Landau level. Some of this work has focused on the nature
of the quantized Hall states, in which the Hall resistance $\rho_{xy}$ is
locked to a rational multiple of the unit $h/e^2$, while the longitudinal
resistivity $\rho_{xx}$ is found to vanish in the limit of low temperatures.
Other work, at least equally interesting, has focused on
Landau-level filling-fractions $f$ where the
quantized Hall effect is not observed. A number of remarkable phenomena have
been observed under these conditions, and we find that the effects of
electron-electron interactions in a two-dimensional electron system in a strong
magnetic field, can be quite subtle, even when the Hall conductance is not
quantized. It seems appropriate to denote these phenomena, collectively, as the
unquantized quantum Hall effect (UQHE).

In the following sections, I will try to outline developments in several
selected areas that I have found to be particularly exciting or challenging.
The references cited are representative examples intended to give the reader
an entrance into the literature of these subjects. It would not be possible
in the space allotted to give anything approaching a complete listing of the
significant contributions, nor has it been possible for me to devote the
time that would be necessary to assemble such a listing,  or to conduct the
thorough critical evaluation that might allow for an optimal use of the
reference space that is available. I apologize in advance for the
references omitted.
\vskip14truept
\noindent
{\bf 2. Fermion Chern-Simons Theory}
\vskip14truept
The unquantized Hall effect has been especially interesting for very high
mobility samples at or near an even-denominator fraction such as
$f=1/2$. The theoretical approach which has proved most useful in this case
is based on a mathematical transformation to a system of fermions (``composite
fermions'') interacting with a fictitious gauge field of the Chern-Simons
type.$^{1-3}$
(Composite fermions were originally introduced by Jain to explain the most
prominent fractional quantized Hall states, at fractions of the form
$f=p/(2p+1)$, where $p$ is an integer.$^{4-6}$ Theories based on a
transformation
to a system of {\it bosons} interacting with a Chern Simons field had also
been used previously to discuss the quantized Hall effect.$^{6,7}$)
A key result has been the realization that at $f=1/2$, and at
various other even denominator filling fractions, there are fermionic
quasiparticles
which move in straight lines, and behave in many ways like quasiparticles of
a Fermi liquid in zero magnetic field.$^1$ For magnetic fields which deviate
by a small amount $\Delta B$ from the magnetic field $B_{1/2}$
which corresponds to $f=1/2$, the quasiparticles move in circles whose
radius $R_c^*$ is equal to the cyclotron radius for a particle in the
{\it effective field} $|\Delta B|$.  This has important experimental
consequences, which have been beautifully demonstrated by several
experiments during the past year.$^{8-10}$

Later in this session, Bob Willett will describe surface acoustic wave
experiments which show a resonance feature when the effective cyclotron
diameter
$2R_c^*$ coincides with a theoretically-predicted multiple of the acoustic
wavelength.$^8$ Other experiments have seen effects in
transport properties when $2R_c^*$ is related to
geometric features of an imposed lithographic structure.$^{9,10}$ Taken
together,
these experiments confirm that fermionic quasiparticles exist and follow the
prescribed trajectories at least  over distances of several microns,
a hundred times larger than the electron-electron separation or the true
cyclotron diameter of an electron in the lowest Landau level.

An important open question is whether some type of modified Fermi liquid
theory can hold, in principle, all the way down to zero energy and infinite
length scales, at zero temperature, precisely at $f=1/2$, in an ideal sample
with no impurity scattering. If so, is there a finite effective mass for the
fermions, or is the effective mass $m^*$ singular in the limit $|E-E_F|\to0$?
The original analysis of Ref.~1 suggested that $m^*$ should diverge as
$\ln|E-E_F|$, if the electron-electron repulsion has the Coulomb form,
$\propto 1/r$ for large separations $r$. More recent investigations do not
necessarily agree with this conclusion, however.$^{11}$ It is also possible
that
different definitions could lead to different results for the effective mass.
One choice, which is at least well defined in principle, is to use the energy
gap $E_g$ of the fractional quantized Hall state at $f=p/(2p+1)$, where $p$
is an integer, to define the mass at energy scale $E_g$, through the
asymptotic relation (presumed valid at large $p$)$^1$
$$
E_g\sim {B\;e\;\hbar\over(2p+1)\;c\;m^*(E_g)}\;.
\eqno(1)
$$
It is difficult to apply this expression to actual experimental data for the
energy gap, however, because of the large effects of impurity
scattering,$^{12}$ for
which there is no proper theory. Values of the effective mass have also been
obtained recently from the amplitude of Shubnikov-de~Haas oscillations at
higher temperatures; however, these results are obtained using a theory of
Shubnikov-de~Haas oscillations based on ordinary Fermi liquid theory which
may or may not be correct near $f=1/2$ for the temperatures in question.$^{13}$

Within the composite fermion  picture, at the mean field level,
the ground state of the fractional quantized Hall state at $f=p/(2p+1)$ is just
an integer quantized Hall state, with $|p|$ filled Landau levels, as
originally noted by Jain. The effective
field $\Delta B$ is equal to $B/(2p+1)$ in this case, and the gap (1) is
identified with the cyclotron energy of a particle of mass $m^*$ in the
field $\Delta B$. In order to properly obtain the dispersion relation for
neutral excitations (quasi-exciton modes) or to study the linear response
functions at finite $q$ and $\omega$, it is necessary to go beyond the mean
field
approximation, at least to the level of random phase approximation, or, better,
to modifications of   the RPA based on Landau Fermi-liquid theory.$^{1,6,14}$
\vskip14truept
\noindent
{\bf 3. Effects of Disorder}
\vskip14truept
A different aspect of the unquantized Hall effect is the transition from one
quantized Hall plateau to another under conditions where impurity scattering
plays a dominant role. When impurities are important we must distinguish more
carefully between the filling fraction $f$, which is defined in
terms of the electron-density $n_e$ by
$$
f={n_ehc\over Be}\;,
\eqno(2)
$$
and the dimensionless Hall conductance $\nu$, which is defined by
$$
\sigma_{xy}={\nu e^2\over h}\;.
\eqno(3)
$$

If the impurity scattering is sufficiently strong, relative to the
electron-electron interaction, there should be a {\it direct transition} from
one
integer Hall plateau to another, in an
{\it interval of magnetic fields} (or of $f$) that becomes
{\it infinitely narrow} in limit of temperature $T\rightarrow 0$.
If the impurity scattering is reduced, then the first fractional Hall
plateau appears, say at $\nu=1/3$.  In this case, according to recent
theoretical analyses,$^{15}$
we  should find a
reentrant phase diagram, where in the limit of
$T\rightarrow 0$, as the magnetic field decreases and $f$ increases,
we find a sequence of sharp transitions:
first from an insulating state with $\nu=0$ to a fractional quantized
Hall state with $\nu=1/3$,
then back down to $\nu=0$, followed by a sharp transition to $\nu=1$. (A direct
transition from $\nu=1/3$ to $\nu=1$ is not allowed.) If the
impurity scattering is decreased further, we should expect to see new plateaus
appear at $\nu=2/3$, $\nu=2/5$, etc. If the impurity scattering is reduced
sufficiently, we reach the situation where at least for the lowest attainable
temperatures, there is a smooth variation of the Hall conductance near
$f=1/2$, and the phenomena characteristic of an impurity-free system begin to
appear.

Recent theoretical work has dealt with such questions as: Which transitions
are allowed to be direct at $T=0$? What are the transition widths at $T\not=0$?
What is the value of $\rho_{xx}$ at the peak of the transition?$^{15-17}$ Other
work has
focused on the  conductivity at $T\not=0$ in the field region of a quantized
Hall plateau, and on the possibility that mesoscopic density
inhomogeneities may be responsible for the observed values of $\rho_{xx}$
at higher temperatures, where Hall plateaus have disappeared.$^{18,19}$
\vskip14truept
\noindent
{\bf 4. Edge States}
\vskip14truept
If a sample is in a quantized Hall state, so that there is an energy gap for
delocalized excitations in the ``bulk'' of the sample, there must nevertheless
be
zero-energy excitations (edge states) along the sample boundaries.  For
noninteracting electrons, in an integer quantized Hall state, the edge states
may
be understood as arising because the occupied Landau levels are pushed up
through the Fermi level by the confining potential at the boundary. (See
Fig.~1,
left panel). Electrons in edge states at a given boundary have a group
velocity,
parallel to the boundary, in a single direction, arising from $\vec E\times\vec
B$
drift of the orbits in the electric field of the confining potential.$^{20}$ An
alternate
view is to think about each edge state as the dividing line between two
incompressible quantum Hall fluids with different values of $\nu$.
An extra particle at the edge causes a bulge in the boundary, which propagates
with a well-defined velocity and direction as an ``edge
magnetoplasmon''$^{21-23}$.
(See right panel of Fig.~1).

Recent theories have extended these considerations to fractional quantized Hall
states, and have characterized the possible combinations of edge states that
may occur.$^{24-27}$ In the fractional case, quantum
fluctuations have a major effect,
particularly on the tunneling of charged quasiparticles
into or out of an edge state. The effects of these fluctuations can be
understood  using the various techniques previously applied to conventional
one-dimensional metals, and the resulting description of the edge is
characterized as a ``chiral Luttinger liquid.''$^{22,27}$
Recent tunneling experiments confirm key predictions of the
theory.$^{28}$

In most actual samples, the electron profile drops gradually to zero at the
edge, on a scale large compared to the electron-electron separation. In a
number
of recent papers, the authors have attempted to calculate self-consistently
the electron profile near a sample edge, and have discussed some of the
differences
that may occur between properties of gradual and  sharp
edges.$^{23,29}$
\vskip14truept
\noindent
{\bf 5. Bilayer Systems}
\vskip14truept
A variety of interesting theoretical and experimental questions arise when
there are two parallel electron layers.
(A wide single quantum
well may also act like a two layer system because the self-consistent Coulomb
potential develops a peak at the center of the well.)
Depending on the separation
between the layers, the Coulomb interaction between electrons in different
layers may be relatively weak or may become comparable in strength to the
interaction between electrons in the same layer. Depending on the height as
well
as the thickness of the barrier between the layers, tunneling of electrons
between the two layers may be more or less important.
Depending on these
parameters, and on the filling factors of the layers,  a variety
of quantized Hall states, as well as unquantized states,  have been found
to occur.$^{30-33}$

One interesting experimental result has been the observation of a quantized
Hall plateau at total filling $f=1/2$ for certain ranges of the system
parameters.$^{30-32}$  Another interesting phase can occur at total fillings
$f=1$,
(and at various other filling factors), where the total filling is locked
at a quantized value, but the relative fraction of electrons in each layer
is free to vary. Interest has focussed on characterizing the possible phases
that can occur, and understanding their properties, as well as on predicting
the occurrence of transitions between different phases as the system
parameters are varied at a fixed filling factor $f$.$^{30-36}$

Application of a magnetic field parallel to the surface introduces,
effectively,
a spatial variation in the phase of the tunneling matrix element between the
two layers. An unexpected phase transition, observed in a bilayer system with
$f=1$, in a relatively weak parallel field, has been explained in terms of this
effect.$^{33,37}$
\vskip14truept
\noindent
{\bf 6. Other Topics}
\vskip14truept
Theoretical progress has also been made on a variety of other aspects of
two-dimensional electron systems in strong magnetic fields. I cite but a
few examples.

In any given sample, for {\it sufficiently strong magnetic fields}, at low
enough
temperatures, one expects to find an {\it insulating} phase where $\rho_{xx}$
becomes
very large, and $\rho_{xy}$ is small compared to $\rho_{xx}$. This could occur
because of electron-electron interaction (formulation of a Wigner crystal),
because of electron impurity interactions (carrier freeze-out) or because
of some complicated combination of these effects. There has been  considerable
theoretical and experimental effort concerning the transition to the
insulating state under various circumstances.$^{38}$

Theoretical arguments have been advanced to explain the observations of a
pseudogap for tunneling into a layer (or between two layers) in the
unquantized Hall regime.$^{39}$

Electron-electron interactions have been invoked to explain the splittings
and shifts of the cyclotron resonance line at very low filling factors.$^{40}$

Invited papers in other sessions of this conference will describe experiments
in these and other areas of the quantum Hall effect, and no doubt they will
also give further references to theoretical work in the field.
\vskip14truept
\noindent
{\bf 7. Acknowledgments}
\vskip14truept
I have benefited from discussions with many workers in the field, but I would
like to thank particularly R.L.~Willett, J.~Eisenstein and B.~Shklovskii,
as well as my direct collaborators P.A.~Lee, N.~Read, S.H.~Simon, S.~He
and P.M.~Platzman. This work has been supported in part by NSF grant
DMR~91--15491.
\vskip14truept
\noindent
{\bf 8. References}
\vskip14truept
\item{1.} B.I. Halperin, P.A. Lee, and N. Read, {\it Phys. Rev. \bf B47},
7312 (1993).
\item{2.} V. Kalmeyer and S.-C. Zhang, {\it Phys. Rev. \bf B46}, 9889 (1992);
E. Rozayi and N. Read, {\it Phys. Rev. Lett. \bf 72}, 900 (1994).
\item{3.} See also G. Moore and N. Read, {\it Nucl. Phys. \bf B360}, 362
(1991);
M. Greiter and F. Wilczek, {\it Mod. Phys. Lett. \bf B4}, 1063 (1990).
\item{4.} J.K. Jain, {\it Phys. Rev. Lett. \bf 63}, 199 (1989); {\it Phys. Rev.
\bf 40}, 8079 (1989); {\it Phys. Rev. \bf B41}, 7653 (1990); J.K. Jain, S.A.
Kivelson, and N. Trivedi, {\it Phys. Rev. Lett. \bf 64}, 1297 (1990); X.G. Wu,
G. Dev
and J.K. Jain, {\it Phys. Rev. Lett. \bf 71}, 153 (1994).
\item{5.} A. Lopez and E. Fradkin, {\it Phys. Rev. \bf 44}, 5246 (1991).
\item{6.} E. Fradkin, {\it Field Theories of Condensed Matter Systems}
(Addison-Wesley,
Redwood City, CA, 1991), Chap.~10.
\item{7.} S.M. Girvin and A.H. MacDonald, {\it Phys. Rev. Lett. \bf 58}, 1252
(1987);
S.-C. Zhang, H. Hansson, and S. Kivelson, {\it Phys. Rev. Lett. \bf 62}, 82
(1989);
{\it Phys. Rev. Lett. \bf 62}, 980(E), (1989); S.-C. Zhang, {\it Int. J. Mod.
Phys.
\bf B6}, 25 (1992); D.-H. Lee and M.P.A. Fisher, {\it Phys. Rev. Lett. \bf 63},
903 (1989).
\item{8.} R.L. Willett, R.R. Ruel, K.W. West, and L.N. Pfeiffer,
{\it Phys. Rev. Lett. \bf 71}, 3846 (1993).
\item{9.} W. Kang, H.L. Stormer, L.N. Pfeiffer, K.W. Baldwin, and K.W. West,
{\it Phys. Rev. Lett. \bf 71}, 3850 (1993).
\item{10.} V.J. Goldman, B. Su, and J.K. Jain, {\it Phys. Rev. Lett. \bf 72},
2065 (1994); V.J. Goldman and B. Su, preprint.
\item{11.} Y.B. Kim, A. Furusaki, X.-G. Wen, and P.A. Lee, preprint; B.L.
Altshuler, A. Houghton, and J.B. Marston, preprint; J. Gan and E. Wong,
{\it Phys. Rev. Lett. \bf 71}, 4226 (1993); D.V. Kveshchenko and P.C.E. Stamp,
{\it Phys. Rev. Lett. \bf 71}, 2118 (1993); C. Nayak and F. Wilczek, {\it Nucl.
Phys. \bf B417}, 359 (1994); J. Polchinski, preprint.
\item{12.} R.R. Du, H.L. Stormer, D.C. Tsui, L.N. Pfeiffer, and K.W. West,
{\it Phys. Rev. Lett. \bf 70}, 2944 (1993).
\item{13.} D.R. Leadley, R.J. Nicholas, C.T. Foxon, and J.J. Harris,
{\it Phys. Rev. Lett. \bf 72}, 1906 (1994); R.R. Du, H.L. Stormer, D.C. Tsui,
L.N. Pfeiffer, and K.W. West, {\it Solid State Commun. \bf 90}, 71 (1994).
\item{14.} A. Lopez and E. Fradkin, {\it Phys. Rev. \bf B47}, 7080 (1993);
S.H. Simon and B.I. Halperin, {\it Phys. Rev. \bf B48}, 17368 (1993); S. He,
S.H. Simon, and B.I. Halperin, {\it Phys. Rev. \bf B50}, 1823 (1993).
\item{15.} S. Kivelson, D.-H. Lee, and S.-C. Zhang, {\it Phys. Rev. \bf B46},
2223 (1992).
\item{16.} D.B. Chklovskii and P.A. Lee, {\it Phys. Rev. \bf B48}, 18060
(1993);
D.-H. Lee, Z. Wang, and S. Kivelson, {\it Phys. Rev. Lett. \bf 70}, 4130
(1993); A.M. Dykhne and I.M. Ruzin, {\it Phys. Rev. \bf B50}, 2369 (1994);
I.M. Ruzin and S. Feng, preprint.
\item{17.} H. Levine, S. Libby, and A. Pruisken, {\it Phys. Rev. Lett. \bf 51},
1915 (1983); J.T. Chalker and P.D. Coddington, {\it J. Phys. \bf C21}, 2665
(1988); Y. Huo and R.N. Bhatt, {\it Phys. Rev. Lett. \bf 68}, 1375 (1992);
A.W.W. Ludwig, M.P.A. Fisher, R. Shantar, and G. Grinstein, preprint.
\item{18.} D.G. Polyakov and B.I. Shklovskii, {\it Phys. Rev. \bf B48}, 1167
(1993),
and preprints
\item{19.} I.M. Ruzin, {\it Phys. Rev. \bf B47}, 15727 (1993); S.H. Simon
and B.I. Halperin, preprint; J. Hajdu, M. Metzler, and H. Moraal, preprint;
M.B. Isichenko, {\it Rev. Mod. Phys. \bf 64}, 961 (1992).
\item{20.} B.I. Halperin, {\it Phys. Rev. \bf B25}, 2185 (1982).
\item{21.} D.B. Mast, A.J. Dahm, and A.L. Fetter, {\it Phys. Rev. Lett. \bf
54},
1706 (1985); D.C. Glattli, E.Y. Andrei, G. Deville, J. Portrenaud, and
F.B.I. Williams, {\it Phys. Rev. Lett. \bf 54}, 1710 (1985).
\item{22.} X.-G. Wen, {\it Phys. Rev. \bf B44}, 5708 (1991).
\item{23.} I.L. Aleiner and L.I. Glazman, preprint.
\item{24.} A.H. MacDonald, {\it Phys. Rev. Lett. \bf 64}, 222 (1990); M.D.
Johnson
and A.H. MacDonald, {\it Phys. Rev. Lett. \bf 67}, 2060 (1991).
\item{25.} A. Cappelli, C.A. Trugenberger, and G.R. Zembla, {\it Nucl. Phys.
\bf B396}, 465 (1993), and preprint.
\item{26.} L. Brey, preprint; D.B. Chklovskii, preprint.
\item{27.} C. de C. Chamon and X.-G. Wen, {\it Phys. Rev. Lett. \bf 70}, 2605
(1993);
K. Moon, H. Yi, C.L. Kane, S.M. Girvin, and M.P.A. Fisher, {\it ibid.
\bf 71}, 4381 (1993); C.L. Kane, M.P.A. Fisher, and J. Polchinski, {\it ibid.
\bf 72}, 4129 (1994).
\item{28.} F.P. Milliken, C.P. Umbach, and R.A. Webb, preprint.
\item{29.} D.B. Chklovskii,
B.I. Shklovskii, and L.I. Glazman, {\it Phys. Rev. \bf B46}, 4026 (1992);
C.W.J. Beenaker, {\it Phys. Rev. Lett. \bf 64}, 216 (1990); A.M. Chang,
{\it Solid State Commun. \bf 74}, 871 (1990); J. Dempsey, B.Y. Gelfand, and
B.I. Halperin, {\it Phys. Rev. Lett. \bf 70}, 3639 (1993); C. de C. Chamon
and X.-G. Wen, preprint.
\item{30.} Y.W. Suen, L.W. Engel, M.B. Santos, M. Shayegan, and D.C. Tsui,
{\it Phys. Rev. Lett. \bf 68}, 1379 (1992).
\item{31.} J.P. Eisenstein, G.S. Boebinger, L.N. Pfeiffer, K.W. West, and S.
He,
{\it Phys. Rev. Lett. \bf 68}, 1383 (1992).
\item{32.} Y.W. Suen, H.C. Manoharan, X. Ying, M.B. Santos, and M. Shayegan,
{\it Surface Science \bf (305}, 13 (1994).
\item{33.} S.Q. Murphy, J.P. Eisenstein, G.S. Boebinger, L.N. Pfeiffer, and
K.W. West, {\it Phys. Rev. Lett. \bf 72}, 728 (1994).
\item{34.} T. Chakraborty and P. Pietil\"ainen, {\it Phys. Rev. Lett. \bf 59},
2784 (1987); S. He, S. Das Sarma, and X.C. Xie, {\it Phys. Rev. \bf B47},
4394 (1993); M. Greiter, X.-G. Wen, and F. Wilczek, {\it Phys. Rev. \bf B46},
9586 (1992); B.I. Halperin, {\it Surface Science \bf 305}, 1 (1994).
\item{35.} X.-G. Wen and A. Zee, {\it Phys. Rev. \bf B47}, 2265 (1993);
J. Fr\"ohlich and A. Zee, {\it Nucl. Phys. \bf 364B}, 517 (1991).
\item{36.} D. Schmeltzer and J.L. Birman, {\it Phys. Rev. \bf B47}, 10939
(1993);
Z.F. Ezawa and A. Iwazaki, {\it Int. J. Mod. Phys. \bf B19}, 3205 (1992);
{\it Phys. Rev. \bf B47}, 7295 (1993).
\item{37.} K. Yang, K. Moon, A.H. MacDonald, S.M. Girvin, D. Yoshioka, and
S.-C. Zhang, {\it Phys. Rev. Lett. \bf 72}, 732 (1994).
\item{38.} See, e.g., R. Price, P.M. Platzman, S. He, and X. Zhu, {\it Surface
Science \bf 305}, 126 (1994); X. Zhu and S.G. Louie, {\it Phys. Rev. Lett. \bf
70},
335 (1993); K.S. Esfarjani, S.T. Chui, and X. Qiu, {\it Phys. Rev. \bf B46},
4638 (1992); L. Ziang and H. Fertig, {\it Phys. Rev. \bf B50} (in press).
\item{39.} S.R.E. Yang and A.H. MacDonald, {\it Phys. Rev. Lett. \bf 70}, 4110
(1993);
Y. Hatsugai, P.A. Bares, and X.-G. Wen, {\it Phys. Rev. Lett. \bf 71}, 424
(1993);
S. He, P.M. Platzman, and B.I. Halperin, {\it Phys. Rev. Lett. \bf 71}, 777
(1993);
P. Johansson and J.M. Kinaret, {\it Phys. Rev. Lett. \bf 71}, 1435 (1993); A.L.
Efros
and F.G. Pikus, {\it Phys. Rev. \bf B48}, 14694 (1993); Y.B. Kim and X.-G. Wen,
preprint; I.L. Aleiner, H.U. Baranger, and L.I. Glazman, preprint.
\item{40.} N.R. Cooper and J.T. Chalker, {\it Phys. Rev. Lett. \bf 72}, 2057
(1994).

\vfill
\eject

\noindent
\medskip
{
\noindent
{\bf Figure 1.} Alternate descriptions of edge states, for noninteracting
spinless
electrons, with two filled Landau levels. The left panel shows the
single-electron
energy spectrum, $\epsilon_{kn}$, in the Landau gauge, near a sample boundary.
The wave function with wavevector $k$ in the $x$ direction is localized about
$y=y_k\equiv k\hbar c/Be$. Heavy lines indicate occupied states; arrows point
to
edge states at the Fermi level. The right panel shows regions of space where
there are respectively 2, 1 and 0 occupied Landau levels. An electron added to
the
inner edge state is here indicated as a bulge, which propagates to the right,
as
indicated. Electron-electron interactions modify the velocities, but not the
directions of propagation.
\bigskip
}

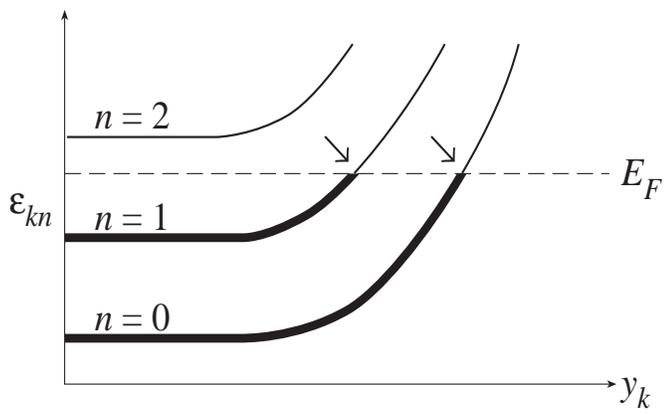
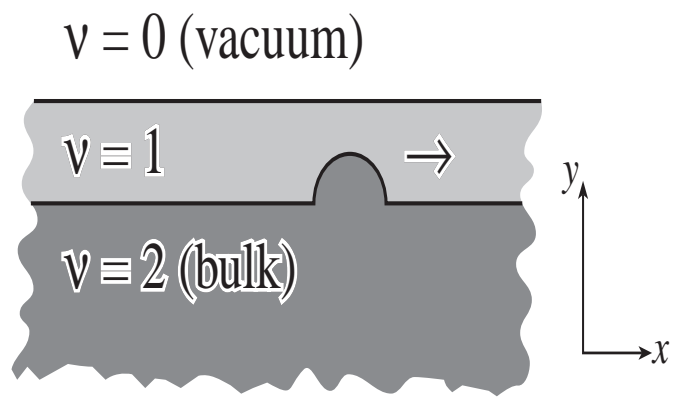
\bye